# Laminar mixed convection in an eccentric annular horizontal duct for a thermodependent non-Newtonian fluid

N. Ait Messaoudene[1*], A. Horimek[1], C. Nouar[2] and B. Benaouda-Zouaoui[1]

[1] Laboratoire des applications énergétiques de l'Hydrogène (LApEH), Saad-Dahlab University of Blida, Algeria

[2] LEMTA, UMR 7563, CNRA – Nancy Université, 2 Avenue de la Forêt de Haye, 54504 Vandoeuvre-les-Nancy, France

[*] Presently at the Department of Mechanical Engineering, Faculty of Engineering, University of Hail, Saudi Arabia

[1*]Corresponding author: N. Ait Messaoudene (full last name : Ait Messaoudene);
naitmessaoudene@yahoo.com;
Laboratoire des applications énergétiques de l'Hydrogène (LApEH),
Université Saad-Dahlab de Blida, BP 270, Blida 09000, Algérie ;
tel. : (231) 557 246 950 ; fax : (213) 25 433 632

**Abstract:**

The present work focuses on the study of mixed convection of a purely viscous shear-thinning fluid in a horizontal annular eccentric duct. The inner and outer cylinders are heated with constant and uniform heat flux densities. The objective of this work is to study the effect of the variation of eccentricity, rheological behavior of the fluid as well as the thermo-dependency of the rheological parameters on the reorganization of the flow and thermal stratification caused by the buoyancy forces. At the entrance of the heating zone, the dynamic regime is assumed to be established and the temperature profile uniform. The conservation equations are solved numerically using a finite difference method with implicit schemes. A secondary azimuthal flow, induced by natural convection, develops downstream of the inlet section. This flow creates a stratification of the thermal field on a given section of the duct, which intensifies downstream from the entrance. On the other hand, the decrease in consistency with increasing temperature near the heated walls produces a centrifugal radial flow towards the walls. The presence of an eccentricity induces in turn a significant effect on the main dynamic field and the stratification of the thermal field. Two cases of upward and downward eccentricity are treated. These show that an upward shift increases the stratification of the thermal field, while the stratification begins to weaken from a certain amount of eccentricity in the case of downward shift. This represents an important result in terms of possible industrial applications. We may indeed conclude that an appropriate choice of downward eccentricity can reduce the thermal stratification, observed experimentally in the case of a concentric heated annular duct [1], when this stratification is undesirable. The choice of this eccentricity depends on rheological and thermal properties of the fluid.

**Keywords:** Shear-thinning fluid, Annular geometry, Eccentricity, Mixed convection, Thermodependency.

**List of Symbols:**

*a*: Location of the positive pole *p* of the bipolar system (*m*)

*B* : Thermal expansion Coefficient ($°C^{-1}$)

Br : Brinkman number $= \mu_0 U_d^2 (1-r_1)^2 / \Phi R_2$

$C_p$: Specific heat ($J.Kg^{-1}.°C^{-1}$)

Gr : Grashof number $= \rho_e^2 g B \Delta T R_2^3 / \mu_0^2$

*h*: Coordinate scale factor (*m*)

*K* : Fluid consistency ($Pa.s^n$)

*n* : Shear-thinning index

Nu : Nusselt number $= 2(1-r_1)/(\theta-\theta_m)$

Pe : Péclet number $=$ Re.Pr

Pn : Pearson number $= b \Phi R_2 / \lambda$

Pr : Prandtl number $= \mu_0 C_P / \lambda$

Re : Reynolds number $= \rho_e U_d R_2 / \mu_0$

$R_1$: Inner cylinder radius (*m*)

$R_2$: Outer cylinder radius (*m*)

$r_1$: Radius ratio $= R_1/R_2$

$T_e$ : Inlet temperature (°C)

*U*: Dimensionless radial velocity

$U_d$: Mean axial velocity ($m.s^{-1}$)

*V*: Dimensionless azimuthal velocity

*W*: Dimensionless axial velocity

*z* : Dimensionless axial coordinate

$z^*$ : Axial coordinate (*m*)

Greek symbols

$\alpha$ : Radial bipolar coordinates

$\beta$ : Azimuthal bipolar coordinate

$\rho$ : Fluid density ($Kg.m^{-3}$)

$\lambda$ : Thermal conductivity ($W.m^{-1}.°C^{-1}$)

$\varepsilon$ : Dimensionless eccentricity

$\theta$ : Dimensionless temperature

$\theta_m$: Mean dimensionless temperature $= \dfrac{1}{SW_m} . \int\limits_{0}^{\pi} \int\limits_{\alpha_1}^{\alpha_2} W \theta h^2 d\alpha . d\beta$

$\Phi$ : Heat flux density ($W.m^{-2}$)

$\Delta T$ : Temperature difference $= \Phi (R_2 - R_1) / \lambda$

$\bar{\mu}_a$ : Dimensionless Apparent viscosity

$\mu_0$ : Wall-shear viscosity at the entrance section ($Pa.s$)

Subscript

1,2 : Cylinders index, (1:inner, 2:outer)

$fd$ : Fully developed

## 1. Introduction

Thermization of non-Newtonian fluids is of great importance in many industries such as papermaking, food and chemical processes. This thermization is generally performed in tubular heat exchangers. Knowing the distribution of the dynamic and thermal fields of the fluid is essential for an optimal control of the thermization process. This depends on the geometrical parameters of the exchanger, the dynamical parameters of the flow, the physical characteristics of the fluid and its rheological behavior.

In certain industries, heating is conducted in horizontal annular ducts [2]. The heated fluids are typically non-Newtonian, with high Prandtl number values and temperature dependent density and consistency. The presence of natural convection generates a secondary azimuthal flow that induces a thermal stratification within a given flow section. Another secondary radial flow arises starting from the inlet section, due to the change of consistency with temperature, and in turn contributes to the phenomenon of thermal stratification.

Thermization problems in annular geometry with two concentric cylinders have been extensively studied. For the case of mixed convection, Kotake *et al.* [3] examined the problem using a Newtonian fluid; two thermal conditions of Neumann type have been selected. Nazrul *et al.* [4] treated the problem for the case of heating air and water, with constant heat flux density applied on the inner cylinder and an adiabatic outer cylinder. The effects of rheological behavior and thermodependency were very well exposed in the study of Nouar *et al.* [1]; the fluid investigated was a pseudoplastic fluid with temperature dependent viscosity. Boundary conditions of Neumann type were considered, with constant heat flux densities applied on both cylinders. Many results have been presented and the authors have very clearly shown, numerically and experimentally, that a secondary azimuthal flow develops starting from the inlet section owing to the effect of buoyancy forces. This flow becomes the dominant heat transfer mechanism starting from a critical axial position, causing the stratification of the thermal field between the top and bottom of the duct which increases downstream. In their experiments, the temperature difference between the two parts can well exceed 20°C at a moderate distance from the entrance of the heating zone (see Fig. 14 (b) p447 in [1]).

Thus, the fluid in the upper (warmer) part could reach a high enough temperature to be denaturized, which is highly undesirable in many cases such as food processing industries. From these results and in order to solve this problem, the application of an eccentricity between the two cylinders (upward or downward) is proposed as a solution. The objective is to seek a proper eccentricity choice for minimizing thermal stratification depending on the rheolgical, geometrical and dynamical parameters.

For heat exchangers of eccentric annular geometry, the forced convection case is analyzed by many authors. Feldman *et al.* [5] have treated the problem of the dynamic regime establishment in a horizontal eccentric annular duct; the Navier-Stokes and continuity equations are solved using bipolar coordinates system. Their results give the required establishment length for different eccentricities and geometrical ratios, as well as the variation in the pressure gradient and axial velocity along the entrance zone. Fang *et al.* [6] have examined the effects of eccentricity on the distribution of axial velocity, pressure gradient and friction factor for a fully developed flow of a shear-thinning fluid; no heating of the walls is considered. Manglik *et al.* [7] have treated the same problem presented in [6], but for a Newtonian fluid with two thermal conditions (imposed flux and imposed temperature) on the inner cylinder, the outer cylinder being considered adiabatic. The same authors have extended their study to the case of a shear-thickening fluid [8]. Feldman *et al.* [9] have studied the problem of the thermal entrance region for the case of an incompressible Newtonian fluid; the dynamic regime is supposed fully developed and the temperature uniform at the entry, different thermal conditions being considered. When viscous dissipation is taken into account as it is encountered in polymers (non-Newtonian fluids) melting manufacturing, annular eccentric geometry is largely present, because the die mandrel is displaced downward to get a uniform annular shape at the exit as possible, this kind of problem have been studied numerically and analytically by several authors (see for instance [10,11]). For an eccentric curved annulus, Nobari *et al.* [12] study numerically the problem for a Newtonian fluid using a second order finite difference method. The governing equations are written in the bipolar-toroidal coordinate system. Different eccentricities and curvature radii with four different thermal boundaries are considered. Several results are given with the main one being the enhancement of heat transfer rate for a curved annuli compared to the straight one at large Dean numbers ( $\mathrm{Re}/\sqrt{a.D_h}$ ).

The case of natural convection alone has also benefited of many studies. Ho *et al*. [13] have examined the effects of Prandtl and Rayleigh numbers as well as eccentricity on the temperature distribution and the average Nusselt number between the two cylinders in the case of a Newtonian fluid. El-Shaarawi *et al*. [14] have numerically solved the problem of natural convection inside vertical open-ended annuli with different eccentric ratios. The governing equations are simplified into parabolic ones by neglecting the diffusion term in the main flow direction. Their results show the velocity profile, axial pressure defect, and other heat transfer parameters in detail. Following a different approach, Mokheimer *et al*. [15] have addressed the same problem numerically by solving the equations of motion based on an analytical solution of the energy equation. The main results reflect the induced flow due to buoyancy forces, the temperature field and the evolution of Nusselt number with a presentation of the effects of eccentricity and radius ratios variation. In a different study, the same authors have developed a number of correlations for the same problem using a standard regression technique for determining the maximal induced flow rate, the heat flux absorbed by the fluid and the Nusselt number [16]. Such correlations are of great interest for technical design of heat exchangers when numerical solutions are difficult to implement and experimental results lacking. An experimental study of a similar configuration where the flowing fluid is air is presented by Hosseini *et al*. [17]; a constant heat flux is imposed on the outer cylinder while the inner cylinder is fully insulated. The results show the variation of the heated wall temperature and average Nusselt number along the duct for the case of a single cylinder, two concentric and two non-concentric cylinders for different values of heat flux. The case of an annular geometry with a heated outer cylinder, an insulated inner cylinder and arbitrary eccentricity is treated by Shu *et al*. [18]. The numerical resolution is performed with a differential quadrature method which has the advantage of converging to very accurate results with a simple mesh; the flow structure is presented in the form of isotherms and stream lines with many validations with experimental results.

Nevertheless, the problem of mixed convection in eccentric annular geometries has not yet received enough attention, especially in the case of non-Newtonian fluids. The existence of an eccentricity has a great effect on the profile of the main dynamic field as well as on the stratification of the thermal field caused by the combined effects of natural convection and the variation of viscosity with temperature. As indicated above, our purpose is to investigate the possibility of overcoming the stratification phenomenon by an appropriate choice of eccentricity, according to the rheological and process parameters.

## 2. Description of the problem

A laminar flow of a pseudoplastic fluid in an eccentric horizontal annular duct, where both cylinders are heated with constant heat flux densities, is considered (Fig.1). Changes in the density $\rho$ and consistency $K$ with temperature $T$ are assumed to be described by the following relations $\rho=\rho_e(1-B(T-T_e))$ and $K=A\exp(-bT)$, where $B$ is the thermal expansion coefficient, $\rho_e$ and $T_e$ are the density and temperature at the inlet section. The specific heat $C_p$, thermal conductivity $\lambda$ and shear-thinning index $n$ are assumed constant. In addition, the following assumptions are adopted: (*i*) The Peclet number is large enough ($>10^3$) so that axial diffusion can be neglected in the momentum and energy equations; (*ii*) the Brinkman number is small enough ($<10^{-3}$) for neglecting viscous dissipation; (*iii*) the variation of density is considered only in the buoyancy term (Boussinesq approximation); (*iv*) the radial and azimuthal variation of pressure in a given flow section is weak. Therefore, the pressure is modeled by $P^*=P^*_m(z^*)+P'^*(\alpha,\beta)$, where $P^*_m(z^*)$ is a mean pressure at a given flow section and $P'^*(\alpha,\beta)$ a variable pressure term at the same section. The flow is assumed symmetric about the vertical plane containing the axes of the two cylinders. At the entrance of the heating zone, the dynamic regime is assumed to be fully developed and the temperature profile uniform.

*** Fig.1 ***

## 3. Governing equations

The present work deals with the case of an upward and downward eccentric annular geometry. The solution of the dynamical and thermal fields with a particular attention to the effect of eccentricity on the latter is presented. For the problem formulation, we have used bipolar coordinate system ($\alpha, \beta, z^*$) ([5,10,12,16,19]). This system is related to the Cartesian coordinate system by the following relations:

$$x^* = h^*.\sin\beta \quad ; \quad y^* = h^*.\text{sh}\alpha$$

$$-\infty < \alpha < +\infty \; ; \; 0 \leq \beta \leq 2\pi$$

With: $h^* = a / (\text{ch}\alpha - \cos\beta)$, where $a$ is the location of the positive pole $p$ defined as

$$a = \frac{R_2 - R_1}{2\varepsilon}\left[\left(\frac{R_2 + R_1}{R_2 - R_1} - \varepsilon^2\right)^2 - \frac{4\varepsilon^2 R_2^2}{(R_2 - R_1)^2}\right]^{1/2}$$

$R_1$ and $R_2$ the inner and outer cylinders radii; $\varepsilon$ is the non dimensional eccentricity defined as $\varepsilon = e / (R_2 - R_1)$, with $e$ being the distance between the two cylinder axes; $\beta=0$ always corresponding to the widest part of the cross section by convention (Fig.2).

*** Fig.2 ***

Following the methodology of Gray *et al*. [20], it can be shown that the condition under which the Boussinesq approximation applies is given by $B\Phi R_2 / \lambda \leq 0.1$, a condition which is considered to be satisfied in the present study. The governing equations of the problem can then be written as:

Continuity equation:

$$\frac{1}{h^2}\frac{\partial}{\partial\alpha}(h.U) + \frac{1}{h^2}\frac{\partial}{\partial\beta}(h.V) + \frac{\partial W}{\partial z} = 0 \tag{1}$$

α momentum equation:

$$\frac{1}{\Pr}\frac{1}{(1-r_1)^2}\left[\frac{U}{h}\frac{\partial U}{\partial \alpha}+\frac{V}{h}\frac{\partial U}{\partial \beta}-U.V\sin\beta+V^2\sh\alpha+\sh\alpha_2 W\frac{\partial U}{\partial z}\right]=-\frac{1}{h}\frac{\partial P'}{\partial \alpha}-Gr\Pr.\sh\alpha_2(1-r_1)^2\theta.\cos\beta$$

$$+\frac{1}{\sh\alpha_2}\left(\frac{\partial \bar{\mu}_a}{\partial \alpha}\left[\frac{2}{h^2}\frac{\partial U}{\partial \alpha}-\frac{2.\sin\beta}{h}V\right]+\frac{\partial \bar{\mu}_a}{\partial \beta}\left[\frac{1}{h^2}\frac{\partial U}{\partial \beta}+\frac{1}{h^2}\frac{\partial V}{\partial \alpha}+\frac{\sh\alpha}{h}V+\frac{\sin\beta}{h}U\right]\right)+\frac{1}{h}\frac{\partial \bar{\mu}_a}{\partial z}\frac{\partial W}{\partial \alpha}$$

$$+\frac{1}{\sh\alpha_2}\left(\bar{\mu}_a\left[\frac{1}{h^2}\frac{\partial^2 U}{\partial \alpha^2}+\frac{1}{h^2}\frac{\partial^2 U}{\partial \beta^2}-\frac{2\sin\beta}{h}\frac{\partial V}{\partial \alpha}+\frac{2\sh\alpha}{h}\frac{\partial V}{\partial \beta}-\frac{(\ch\alpha+\cos\beta)}{h}U\right]\right)+O\left(\frac{1}{Pe^2}\right) \qquad (2)$$

β momentum equation:

$$\frac{1}{\Pr}\frac{1}{(1-r_1)^2}\left[\frac{U}{h}\frac{\partial V}{\partial \alpha}+\frac{V}{h}\frac{\partial V}{\partial \beta}+U^2\sin\beta-U.V\sh\alpha+\sh\alpha_2 W\frac{\partial V}{\partial z}\right]=-\frac{1}{h}\frac{\partial P'}{\partial \beta}+Gr\Pr.\sh\alpha_2(1-r_1)^2\theta.\sin\beta$$

$$+\frac{1}{\sh\alpha_2}\left(\frac{\partial \bar{\mu}_a}{\partial \alpha}\left[\frac{1}{h^2}\frac{\partial U}{\partial \beta}+\frac{1}{h^2}\frac{\partial V}{\partial \alpha}+\frac{\sh\alpha}{h}V+\frac{\sin\beta}{h}U\right]+\frac{\partial \bar{\mu}_a}{\partial \beta}\left[\frac{2}{h^2}\frac{\partial V}{\partial \beta}-\frac{2.\sh\alpha}{h}U\right]\right)+\frac{1}{h}\frac{\partial \bar{\mu}_a}{\partial z}\frac{\partial W}{\partial \beta}$$

$$+\frac{1}{\sh\alpha_2}\left(\bar{\mu}_a\left[\frac{1}{h^2}\frac{\partial^2 V}{\partial \alpha^2}+\frac{1}{h^2}\frac{\partial^2 V}{\partial \beta^2}+\frac{2\sin\beta}{h}\frac{\partial U}{\partial \alpha}-\frac{2\sh\alpha}{h}\frac{\partial U}{\partial \beta}-\frac{(\ch\alpha+\cos\beta)}{h}V\right]\right)+O\left(\frac{1}{Pe^2}\right) \qquad (3)$$

z momentum equation:

$$\frac{1}{\sh\alpha_2}\frac{1}{\Pr}\frac{1}{(1-r_1)^2}\left[\frac{U}{h}\frac{\partial W}{\partial \alpha}+\frac{V}{h}\frac{\partial W}{\partial \beta}+\sh\alpha_2 W\frac{\partial W}{\partial z}\right]=-\frac{\partial P_m}{\partial z}+\frac{1}{(\sh\alpha_2)^2}\frac{1}{h^2}\left[\frac{\partial \bar{\mu}_a}{\partial \alpha}\frac{\partial W}{\partial \alpha}+\frac{\partial \bar{\mu}_a}{\partial \beta}\frac{\partial W}{\partial \beta}\right]$$

$$+\bar{\mu}_a\frac{1}{(\sh\alpha_2)^2}\left[\frac{1}{h^2}\frac{\partial^2 W}{\partial \alpha^2}+\frac{1}{h^2}\frac{\partial^2 W}{\partial \beta^2}\right]+O\left(\frac{1}{Pe^2}\right) \qquad (4)$$

Energy equation:

$$\frac{U}{h}\frac{\partial \theta}{\partial \alpha}+\frac{V}{h}\frac{\partial \theta}{\partial \beta}+\sh\alpha_2 W\frac{\partial \theta}{\partial z}=\frac{1}{\sh\alpha_2}(1-r_1)^2\left[\frac{1}{h^2}\frac{\partial^2 \theta}{\partial \alpha^2}+\frac{1}{h^2}\frac{\partial^2 \theta}{\partial \beta^2}\right]+O\left(\frac{1}{Pe^2},Br\right) \qquad (5)$$

Integral continuity equation:

$$\int_0^\pi\int_{\alpha_1}^{\alpha_2}\frac{W}{(\ch\alpha-\cos\beta)^2}d\alpha\,d\beta=\frac{\pi}{2}\left(\frac{1}{(\sh\alpha_2)^2}-\frac{1}{(\sh\alpha_2)^2}\right) \qquad (6)$$

Boundary conditions:

The flow is assumed to be symmetrical with respect to the vertical plane containing the duct axis. In the dimensionless coordinates system, the computational domain is $(\alpha,\beta,z) \in [\alpha_1, \alpha_2]\times[0,\pi]\times[0, z_f]$, where $z_f$ is the non dimensional length of the computational domain.

$$\alpha = \alpha_1, \qquad U = V = W = 0 \;;\qquad \frac{1}{h}\frac{\partial \theta}{\partial \alpha} = -\frac{1}{\sh \alpha_2} \qquad (7a)$$

$$\alpha = \alpha_2, \qquad U = V = W = 0 \;;\qquad \frac{1}{h}\frac{\partial \theta}{\partial \alpha} = +\frac{1}{\sh \alpha_2} \qquad (7b)$$

$$z = 0, \qquad W = U_{fd}(\alpha,\beta); \; U = V = 0; \qquad \theta = 0 \qquad (7c)$$

$$\beta = 0 \wedge \pi, \qquad \frac{\partial U}{\partial \beta} = \frac{\partial W}{\partial \beta} = V = 0; \qquad \frac{\partial \theta}{\partial \beta} = 0 \qquad (7d)$$

Equations (1) to (7) are made dimensionless by using the following scales:

$$z = \frac{z^*}{L}; \quad h = \frac{h^*}{a}; \quad \bar{\mu}_a = \frac{\mu_a^*}{\mu_0}; \quad \theta = \frac{(T-T_e)\lambda}{\Phi R_2};$$

$$P_m = \frac{P_m^* R_2^2}{\mu_o U_d L}; \; P' = \frac{P^{'*} L}{\mu_o U_d}; \; W = \frac{W^*}{U_d}; \; U = \frac{U^* L}{R_2 U_d} = \frac{U^* L}{U_d}\frac{\sh \alpha_2}{a}; \; V = \frac{V^* L}{R_2 U_d} = \frac{V^* L}{U_d}\frac{\sh \alpha_2}{a} \qquad (8)$$

It should be noted that the above mentioned stared quantities are dimensional. $L$ is a typical axial scale length for temperature variations and is given by $L=\rho_e C_P U_d (R_2-R_1)^2/\lambda$. It can be viewed as a length over which downstream convection balances transverse conduction, i.e. $L/R_2(1-r_1)^2=\text{Pe}$ [1,21]. The reference scales for the radial and azimuthal velocities are determined from the continuity equation. This change of variables is adopted in order to have dimensionless variables of order 1.

In the present study, it is assumed that the rheological behavior of the fluid is described by the power law model:

$$\boldsymbol{\tau} = 2\mu_a \boldsymbol{D} \qquad (9)$$

Where $\tau$ and $D$ are the deviatoric extra-stress, and deformation rate tensors. The apparent viscosity $\mu_a$ is given by:

$$\mu_a = K\,(4D_{II})^{(n-1)/2} \qquad (10)$$

Where:

$D_{II} = 1/2 \text{ trace}(\mathbf{D}^2)$

$$= \frac{1}{2}\left\{\left(\frac{1}{h}\frac{\partial U}{\partial \alpha} - \frac{\sin\beta}{a}V\right)^2 + \left(\frac{1}{h}\frac{\partial V}{\partial \beta} - \frac{\text{sh}\,\alpha}{a}U\right)^2 + \left(\frac{\partial W}{\partial z}\right)^2\right\}$$
$$+ \frac{1}{4}\left\{\left(\frac{1}{h}\frac{\partial V}{\partial \alpha} + \frac{1}{h}\frac{\partial U}{\partial \beta} + \frac{U}{a}\sin\beta + \frac{V}{a}\text{sh}\,\alpha\right)^2 + \left(\frac{1}{h}\frac{\partial W}{\partial \alpha} + \frac{\partial U}{\partial z}\right)^2 + \left(\frac{1}{h}\frac{\partial W}{\partial \beta} + \frac{\partial V}{\partial z}\right)^2\right\}$$
(11)

The scalar $D_{II}$ is the second invariant of tensor $\mathbf{D}$. A singularity occurs in equation (10) as $D_{II} \to 0$. To overcome this difficulty, a modified version of the rheological equation is used. When the value of $D_{II}$ is less than a critical value $D_{IIc}$, the apparent viscosity is 'frozen' at a value calculated with $D_{IIc}$. This method was adopted by several authors [1]. The value of $D_{IIc}$ must be sufficiently small in order to describe the rheological behavior of a pseudoplastic fluid, but without creating numerical difficulties, due to too abrupt changes in the apparent viscosity near the point where $D_{II}=0$. In the present work, a value of $D_{IIc} = 2.5 \times 10^{-3}$ is used, for which the results are insensitive to the cut-off value.

## 4. Numerical procedure

The conservation equations with their boundary conditions are discretized by a finite difference technique with an implicit scheme, where partial derivatives in the radial and azimuthal directions are approximated by a centered scheme; those in the axial direction are approximated by a forward upwind scheme. Neglecting axial diffusion makes the problem parabolic and the system of equations (2) to (5) can be solved by a stepwise integration in the axial direction starting from a specified set of upstream initial conditions. The momentum and energy equations are discretized using the two-step alternating direction implicit method (A.D.I). The solving algorithm is based on that proposed by Briley [22]. This algorithm allows treating the three-dimensional parabolic problem as if it were two dimensional. The basic idea is to assume the pressure field at the entrance of the duct to be known (i.e $\partial P_m/\partial z$, $\partial P'/\partial \alpha$ and $\partial P'/\partial \beta$), making it possible to solve the momentum equations separately. Three steps are followed:

-*First step*: Calculate the axial velocity $W^{n+1}$ at section $n+1$ from equation (4) using the axial pressure gradient in the downstream section $(\partial P_m/\partial z)^n$. The calculated axial velocity is used to verify the overall continuity equation (6). The residual flow (Res) defined by equation (12) associated with the secant method is then used to correct the axial pressure gradient. Both processes are repeated until Res $\leq 10^{-7}$. Simpson's rule is used for the calculation of the double integral in (12)

$$\int_0^\pi \int_{\alpha_1}^{\alpha_2} \frac{W}{(\operatorname{ch}\alpha - \cos\beta)^2} d\alpha\, d\beta - \frac{\pi}{2}\left(\frac{1}{(\operatorname{sh}\alpha_2)^2} - \frac{1}{(\operatorname{sh}\alpha_1)^2}\right) = \text{Res} \tag{12}$$

- *Second step*: Calculate velocities $U^{n+1}$ and $V^{n+1}$ from equations (2) and (3) using $(\partial P'/\partial \alpha)^n$ and $(\partial P'/\partial \beta)^n$. The two velocities $U_p$ and $V_p$ (Provisional velocities) which are obtained are corrected to verify the local continuity equation (1):

$$U = U_p + U_c\ ;\qquad V = V_p + V_c \tag{13}$$

It is assumed that the correction velocities ($U_c$, $V_c$) are irrotational, they derive therefore from a potential $\chi$ so that:

$$U_c = \frac{1}{h}\frac{\partial \chi}{\partial \alpha}\ ;\ V_c = \frac{1}{h}\frac{\partial \chi}{\partial \beta} \tag{14}$$

Equations (13) and (14) are substituted into (1); thus we obtain a Poisson equation:

$$\Delta \chi = f \tag{15}$$

Where: $f = -\left(\operatorname{sh}\alpha_2 \frac{\partial W}{\partial z} + \frac{1}{h}\frac{\partial U_p}{\partial \alpha} - \operatorname{sh}\alpha\, U_p + \frac{1}{h}\frac{\partial V_p}{\partial \beta} - \sin\beta\, V_p\right)$

noting that: $\Delta = \left(\frac{1}{h^2}\frac{\partial^2}{\partial \alpha^2} + \frac{1}{h^2}\frac{\partial^2}{\partial \beta^2}\right)$

The method of successive under-relaxation (SUR) with double sweep (following $\alpha$ and $\beta$) is used to solve equation (15). Once the two transverse velocities are corrected, $(\partial P'/\partial \alpha)^{n+1}$ and $(\partial P'/\partial \beta)^{n+1}$ are determined from an equation of Poisson constructed after rearrangement of equations (2) and (3).

$$\Delta P' = S \tag{16}$$

Where: $S = \dfrac{1}{h}\dfrac{\partial \psi_1}{\partial \alpha} - \operatorname{sh}\alpha.\psi_1 + \dfrac{1}{h}\dfrac{\partial \psi_2}{\partial \beta} - \sin\beta.\psi_2$

As for equation (15), the successive under-relaxation method (SUR) with double sweep is used to solve equation (16).

The convergence criterion for $\chi$ and $P'$ are:

$$Max|\chi^{k+1} - \chi^{k}|/Max|\chi^{k+1}| \leq 10^{-5} \qquad Max|P'^{k+1} - P'^{k}|/Max|P'^{k+1}| \leq 10^{-5} \qquad (17)$$

- *Third step*: Solve the energy equation using equation (5) to determine $\theta^{n+1}$.

Because the fluid consistency ($K$) is temperature dependent, the three steps are repeated until convergence.

After several preliminary tests on the dependence of the solution in relation to the mesh size (Fig.3), a regular grid of 101 × 101 points within the radial direction $\alpha$ and the azimuthal direction $\beta$ is adopted (see discussion in the next section). Increasing the eccentricity complicates the problem further, leading to different axial meshes choices. Furthermore, the choice of the axial mesh size is also affected by rheological behavior of the fluid especially for low values of *n* (<0.5); it must be increasingly refined as *n* decreases. In order to overcome these constraints, axial step values $\Delta z = 2.5 \cdot 10^{-4}/L$, $\Delta z = 10^{-4}/L$ and $\Delta z = 10^{-5}/L$ are used.

*** Fig. 3 ***

Furthermore, due to the fact that the transformation from the Cartesian to bipolar coordinate system presents a singularity in the computations if $\varepsilon$ is taken exactly equal to 0 [5,14], a very small value of $\varepsilon$ must be chosen to reproduce the concentric case. Tests performed for $\varepsilon = 10^{-3}$, $10^{-4}$ and $10^{-5}$ show that the last value gives a good representation of the concentric case. Therefore, computations of the concentric case are actually performed for $\varepsilon = 10^{-5}$, noting that $\varepsilon = 0$ is written instead of $\varepsilon = 10^{-5}$ for the sake of simplicity.

Remark: For a better presentation, the results in the radial direction are presented in terms of *r* instead of $\alpha$ ($r_1 \leq r \leq 1.0$). This has no incidence since the radial mesh is regular.

*Validation of the computing code*

At the entrance of the heating zone, the velocity profile is assumed fully developed, this assumption being adopted by several authors [3, 4, 8, 9, 23, 24]. Yet, this profile is strongly affected by the eccentricity $\varepsilon$ between the two cylinder axes. Solving for the fully developed flow in the isothermal case using equations (1) and (4) allows for the determination of the axial velocity profile and the axial pressure gradient $\partial P_m/\partial z$ for each $\varepsilon$ at the heating zone entrance. The results show, as expected, that with increasing eccentricity, the velocity increases in the widest part (Fig. 4.a; $\beta = 0$) and decreases in the narrowest part (Fig. 4.b; $\beta = \pi$) where the flow is almost blocked for eccentricities greater than 0.6. In the case of a Newtonian fluid and for $\varepsilon=0.2$, the maximum velocity in the wide part is 33.3% greater than in the concentric case and 38.7% lower in the narrow part. These values become respectively 54% and 89.7% for an eccentricity of 0.6. For this type of fluid, the results are validated by comparing them with the results of Escudier *et al*. [25] for eccentricity values of 0.2, 0.5 and 0.8 (see [26], p65) and with those obtained by Feldman *et al*. [5] for different eccentricities and different radii ratios (Table 1).

***Table 1***

The shear-thinning index *n* has also a great effect on the axial velocity profile. This is due to increased parietal gradient with decreasing *n*; this increase leads to a decrease of the apparent viscosity of the fluid which leads to greater uniformity of the flow (for this reason, the axial velocity profile becomes flatter as *n* is smaller). A large eccentricity can cause the blockage of the flow in the narrow part, even more so when the shear-thinning index is low. This effect is presented for three values of the shear-thinning index in figure 4. A comparison for $n = 0.6$ and $n = 1.0$ with the results obtained by Manglik *et al.* [7] is also presented in [26] where other validations are made for different situations, including mixed convection, in the concentric case by comparison with the results of Benaouda-Zouaoui [21]. All these show a good agreement with the results obtained by the computational procedure used in the present study and more extensively presented in [26].

***Fig. 4***

*Effect of the mesh size*

In order to test the effect of the transverse mesh ($\alpha \times \beta$) on the results, computations are performed for the case of mixed convection with $\varepsilon=0.2$ and $\varepsilon=0.7$ for five transverse mesh sizes (51×51, 86×86, 101×101, 126×126 and 141×141). The results are compared in terms of the mean interior Nusselt number ($\overline{Nu}_1$) in figure 5. As it can be observed, after just a short distance from the entrance section (z approximately equal to $10^{-5}$; corresponding to a dimensional distance of approximately 3 mm for the presented case with Pe=33440, $R_1$=20 mm and $R_2$=40mm), all the curves for mesh size greater than 51x51 almost coincide. The results become sensitive to the effect of the mesh size when approaching the entrance of the heating zone. This is not surprising, since the thermal boundary layer, is thinner in this area. However, if we accept a variation of 1% at positions greater then z≈3 $10^{-6}$ after the entrance, the grid (101 x 101) is sufficient. Following these considerations, a regular transverse mesh size of 101×101 is finally chosen in the present work since any further refinement does not improve results very much while considerably impeding computing time.

***Fig. 5 ***

## 5. Results and discussion

In order to highlight the effect of eccentricity, that of the rheological behavior and thermodependency of the fluid and that of natural convection on the velocity and temperature profiles, the analysis of the flow is presented in two parts. The first part deals with the forced convection case and the second part is dedicated to the mixed convection case.

All results are presented in nondimensional form, except for the outer cylinder wall temperature variation along the duct length ($T_2$ vs. $z^*$) which is presented dimensionally for the sake of comparison with reference [1] (all computations are made for a computational domain of length $z^*_f$ =3m, $R_2$=40 mm and a radius ratio $r_1$=0.5 for the same purpose). This also allows a better grasping of one of the main objectives of the present work which deals with the temperature difference between the upper and lower parts of the duct. The variation of *Nu* along the heated zone is presented using a logarithmic scale for both *Nu* and *z*, which brings to evidence the axial position from where natural convection becomes the dominant mechanism of heat transfer.

*5.1. Forced convection*

Introducing an eccentricity in the annular duct has a great effect on the profile of the main flow, which results in an acceleration in the wide part and a deceleration in the narrow part. The orientation of the eccentricity (upward or downward) does not matter in this case and all results are linked to the width of the cross section (narrow and wide) whether it is positioned on top or bottom. The acceleration of the flow in the wide part enhances heat exchange in this part. In the narrowest part, the fluid heats up more rapidly, due to the decrease of the fluid stream thickness there. The rapid heating is well noted for eccentricities greater than 0.2, and the heat exchange tends to be more conductive than convective in this part.

Furthermore, for high eccentricities ($\varepsilon > 0.6$), particularly with low *n* values when the axial velocity becomes almost zero in the narrow part of the annular section, the assumption of neglecting axial thermal diffusion can lead to unexpected errors. Feldman *et al*. [9] note that for Peclet numbers < 50, the assumption of neglecting thermal axial diffusion is no longer valid, especially for either low radii ratios or low shear thinning indexes (see also [8], p813). The development of the thermal boundary layer will be much faster in the narrow region than in the wide region (Fig. 6). On the other hand, the temperature difference between top and bottom increases strongly with eccentricity, thus creating a stratification of the thermal field between the two zones (Fig. 7).

\*\*\* Fig. 6 \*\*\*

\*\*\*Fig. 7 \*\*\*

The decrease of consistency *K* with increasing temperature *T* induces a reorganization of the flow. This is characterized by a radial displacement of fluid particles from the central zone toward the heated walls.

The wall velocity gradient increases, thus promoting heat exchange near the walls, whereas the axial velocity decreases in the central zone in order to insure flow rate conservation (Fig. 8.a). This has been well demonstrated by Nouar *et al*. [1] for the concentric case and it is shown here to remain valid for the eccentric case for low eccentricities ($\varepsilon < 0.2$). For high eccentricities, a new phenomenon appears. The flow which was accelerated close to the wall and decelerated in the central zone begins to accelerate in the central zone too; this being observed only in the narrow part. This acceleration is mainly due to the decrease in viscosity because of the strong propagation rate of heat. The acceleration of the whole flow in this part contributes to promote heat exchange and to reduce the problem of flow blockage observed for high eccentricities (Fig. 8.b). The phenomena observed in the wide part of the duct remain unchanged (Fig. 8).

***Fig. 8***

The radial motion of colder fluid particles from the central zone toward the walls tends to cool them and a significantly less pronounced temperature difference between the top and bottom parts of the outer wall is observed compared to the non thermodependent case (Fig. 9).

*** Fig. 9***

The heat exchange coefficient (*Nu*) decreases along the heating zone; this is due to the increase of the bulk fluid temperature. The presence of an eccentricity induces a more rapid increase of the bulk fluid temperature, thus accentuating the reduction of the transfer coefficient. The evolution of the external circumferential average Nusselt number ($\overline{Nu}_2 = 2(1-r_1)/(\overline{\theta}_2 - \theta_m)$) along the duct is presented in Figure 10 for different eccentricities ($\varepsilon=0$; 0.2 and 0.4) and the effect of the eccentricity is very well observed. A similar result is obtained for the internal circumferential average Nusselt number ($\overline{Nu}_1 = 2(1-r_1)/(\overline{\theta}_1 - \theta_m)$) [8].

***Fig. 10 ***

Figure 11 shows an improvement of the Nusselt number which is more and more marked as shear-thinning is important. This improvement is due to an increased axial velocity gradient with decreasing shear-thinning index *n*.

*** Fig. 11***

The acceleration of the main flow near the walls due to the effect of thermodependency leads to improved values of Nusselt number compared to the non thermodependent case (Fig.12). This improvement is even more marked as the Pearson number increases [1].

***Fig. 12 ***

*5.2. Mixed Convection*

This part of the study focuses on the case of mixed convection for a thermodependent shear-thinning fluid. The decrease in fluid density $\rho$ with temperature induces an upward azimuthal flow ($V > 0$) of hot fluid along the walls and a downward flow ($V < 0$) of cold fluid in the central zone. Fluid particles enter the boundary layer in the bottom of the duct and leave it in the upper part (Fig.13).

***Fig. 13 ***

For the case $\varepsilon = 0$ (concentric case), the analysis of the tangential velocity profiles *V* at a given axial position and for different azimuthal positions underlines an azimuthal acceleration of the fluid from the bottom of the duct to the mid plane where $\beta = \pi/2$, then a deceleration toward the top of the duct (Fig.14). This remains valid for the case of upward and downward shift for low eccentricities. For large eccentricities, the plane position of maximum tangential velocity can move beyond $\beta = \pi/2$. The azimuthal flow increases in intensity along the heating zone, and natural convection becomes increasingly the dominant mechanism in heat transfer.

***Fig. 14 ***

The shear-thinning of the fluid ($n\searrow$) induces a decrease of *V*. This evolution can be explained by the fact that shear-thinning induces an increase of the wall axial velocity gradient. Consequently, the thermal boundary layer thickness -of the main flow- decreases, and associated with it, that of the secondary flow. The temperature difference between the wall and the fluid decreases, leading to a decrease in the intensity of recirculation (Fig.15) [1]. This phenomenon is observed for the eccentric cases for all values of *β* and *z*.

*** Fig. 15 ***

The decrease of consistency *K* with temperature reduces sheering stresses, causing an increase in intensity of the recirculating secondary flow. On the other hand, it increases the gradient of the axial velocity near the walls, thus reducing the thickness of the thermal boundary layer and reducing the average wall temperature relative to the non thermodependent case. Figure 16 shows the development of thermal boundary layer for the case of mixed convection for different eccentricity values. In the case of upward shift (Fig. 16.a), the thermal boundary layer develops more rapidly in the upper part of the duct than in the lower part. The profile that was somewhat regular in the case of forced convection (with or without thermodependency) is distorted by the ascending warm streams and downward flow of colder streams. Thermodependency adds more to this distortion because of the radial flow generated by the motion of fluid particles from the mid zone of the flow section toward the walls. In the case of the downward shift (Fig.16.b), and for eccentricities of less than or the order of 0.2, the thermal boundary layer develops more rapidly in the upper part than the lower part, due to the same phenomenon of ascending warm streams. For larger eccentricities, the opposite phenomenon is observed. This is due in part to the closeness of the two cylinders in the bottom part of the annular gap and secondly to the increased thickness of fluid layers in the upper part. We note that for this second case, the development of the thermal boundary layer in the bottom is much slower compared to the first case of upward shift.

***Fig. 16 ***

As in the case of forced convection with $K = f(T)$, the main flow is accelerated near the heated walls due to the decrease in consistency. On the other hand, it is decelerated in the central zone due to the generation of a radial motion of colder fluid particles from the center toward the walls. But because of the secondary flow generated by buoyancy forces, the warmer fluid tends to accumulate in the upper part along the pipe leading to an acceleration of the main flow in this part in order to satisfy continuity (case $\varepsilon=0$). The upward shift increases the fluid temperature in the upper part, leading to an increased acceleration in this part (Fig. 17.a). Comparing with the case of thermodependent forced convection; acceleration in the case of mixed convection is indeed more intense (Fig.8 and Fig.17.a). For a downward shift (Fig. 17.b) and eccentricities of less than or about 0.2, acceleration is observed in the lower part very close to the inlet section when the dominant mechanism in heat transfer is forced convection. When natural convection becomes the dominant mechanism, the reverse phenomenon is observed. The flow slows down progressively in the lower part and accelerates in the upper part due to stratification of the thermal field. For larger eccentricities, a rapid acceleration of the main flow is observed in the lower part near the inlet section. A gradual deceleration in this part is then observed away from the inlet section with the intensification of the secondary flow due to buoyancy forces. In the upper part (wide region), a small reacceleration is encountered far from the inlet section. This is due to the increased thickness of the layers of fluid in this part.

***Fig. 17 ***

The variation of the average external circumferential Nusselt number along the heating zone in the case of an upward shift with $\varepsilon=0.2$ along with the case of a downward shift is presented in Figure 18. Near the inlet section, forced convection is the dominant mechanism in heat transfer, $\overline{Nu_2}$ decreases along the duct with increasing $z$. The effect of natural convection is weak, and all the curves fall on that corresponding to the forced convection. Far from the inlet section, natural convection becomes the dominant mechanism in heat transfer and the curves corresponding to the mixed convection rise above those corresponding to forced convection starting from a certain critical axial position [1]. The downward shift enhances the effect of natural convection. Consequently, the curves corresponding to mixed convection rise above those corresponding to forced convection much closer to the inlet section compared to the concentric case ($\varepsilon=0$) [26]. This phenomenon is further marked with increasing eccentricity. For the upward shift case, the divergence between the two curves (forced and mixed convection) tends to take place farther away from the inlet section compared to the concentric case. This can be explained by the fact that when eccentricity increases, the remoteness of the two cylinders in the lower part (wide part) slows down the development of the thermal boundary layer, consequently weakening the effect of natural convection.

*** Fig. 18 ***

The ascending azimuthal flow which begins right at the entrance of the duct creates a stratification of the thermal field between the top and bottom of the duct. This stratification is observed for both cases of upward or downward shift. In the case of an upward shift, the intensity of the ascending flow decreases with increasing eccentricity because of the increased distance between the two walls in the lower part which is more and more enlarged. However, the closeness of the walls on the top makes the fluid in the top become more heated than the one in the bottom, therefore increasing stratification with increasing eccentricity (Fig.19.a). In the case of a downward shift, a rapid increase in temperature of the outer wall (inner wall too) in the lower part due to the closeness of the two cylinders is observed. This increase is even stronger when eccentricity increases. The ascending flow starts right at the entrance, and a gradual increase in temperature of the outer wall on the top is observed. Far from the entrance, natural convection becomes the dominant mechanism in heat transfer and the temperature of the outer wall on the top becomes higher compared to the bottom. For this case of shift, there are two distinct phenomena. One for small eccentricities ($<\approx 0.2$), where an increase in stratification between the top and bottom of the duct compared to the concentric case is observed. This is mainly due to strong heating of the fluid in the lower part caused by the closeness of the two cylinders associated with the significant effects of natural convection in the top. The second phenomenon is observed for higher eccentricities, where stratification becomes increasingly low when eccentricity increases. In fact, there is a decrease in the mass of the more highly heated fluid in the bottom where the walls are closer. This fluid must in turn exchange heat with more and more important layers of less intensely heated fluid (in the wide part of the duct) during its ascending flow, thus leading to a weaker contribution of the azimuthal flow to the heating process. As a result, a considerable decrease of thermal stratification between the top and bottom parts of the duct can be observed at a moderate distance from the entrance of the heating zone (Fig.19.b).

Moreover, fluid thermodependency (decrease of $K$ with $T$) on the one hand reduces friction forces near the heated walls, which induces an increase in intensity of recirculation of the secondary azimuthal flow. On the other hand, it increases the gradient of the axial velocity near the walls, leading to a decrease of the thermal boundary layer thickness and a decrease of the mean wall temperature relative to the non thermodependent case. This can clearly observed by comparing Fig.19 and Fig.17 (for $r=1$; Top and Bottom).

***Fig. 19 ***

Figure 20 shows the variation of the local external circumferential Nusselt number ($Nu_2=2(1-r_1)/(\theta_2-\theta_m)$) for $\beta=0$ and $\beta=\pi$, for an upward and downward eccentricity ($\varepsilon=0.2$). Away from the inlet section, the secondary flow caused by buoyancy forces becomes increasingly intense. This explains the deterioration of the heat transfer coefficient in the upper part of the annular gap and its improvement in the lower part. Near the inlet section, the difference between $Nu_2(\beta=0)$ and $Nu_2(\beta=\pi)$ is rather small as the dominant mechanism of heat transfer is forced convection.

Far from the inlet section, the difference becomes more marked, thus reflecting the dominant effect of buoyancy forces. This difference is even greater when eccentricity increases for the downward shift case (Fig.20.a). The opposite phenomenon occurs for an upward shift (Fig.20.b). The solid line curves correspond to the non thermodependent case ($Pn=0$) and the dotted ones are for the thermodependent case ($Pn=8$). For both cases presented, $Nu_2$ is higher in the thermodependent case compared to the non thermodependent case. This is due to increased wall gradient of axial velocity and reduced walls temperature (reduction of $\theta_{1,2} - \theta_m$).

***Fig. 20 ***

## 6. Conclusion

The problem of mixed convection for the laminar flow of a shear-thinning fluid with variable consistency in a horizontal eccentric annular duct is solved numerically. The results show that:

- The application of an eccentricity between the cylinders strongly affects the axial velocity by accelerating the flow in the wide part and decelerating it in the narrow part; ultimately almost reaching blockage of the flow in the narrow part particularly for low values of $n$ ($n < 0.5$).
- In the case of forced convection alone, bringing the two cylinders closer in the narrow part enhances the heating of the fluid. This leads to a fast development of the thermal boundary layer, thus creating stratification between the wide and the narrow part of the duct.
- The decrease of $K$ with $T$ close to the hot walls generates a radial motion of cooler fluid particles from the core of the flow section toward the walls, leading to an increase in the wall gradient of axial velocity and thus an acceleration of the fluid near the walls. Subsequently, a deceleration in the core zone occurs in order to ensure the conservation of mass. This is only valid for small eccentricities. For high eccentricities, acceleration is observed across the entire narrow part due to the sharp decrease in viscosity throughout this part. The radial flow contributes to the cooling of the walls and a less pronounced stratification is observed compared to the previous case.
- The decrease of $\rho$ with $T$ for the mixed convection generates an ascending stream of hot fluid near the walls and a downward flowing stream of cooler fluid in the core zone. An azimuthal boundary layer is thus created. The enhancement of the ascending hot streams creates a thermal stratification between the top and the bottom parts of the duct. This stratification is intensified in the case of an upward shift of the inner cylinder, or a downward shift with low eccentricities ($\varepsilon <\approx 0.2$). However, this stratification decreases at a moderate distance from the entrance of the heating zone in the case of a downward shift with eccentricities greater than 0.2.

- For $K = f(T)$, the secondary azimuthal flow generated by the decrease of $\rho$ is enhanced. This is due to the decrease of the viscosity near the walls. A secondary radial flow occurs from the core zone toward the walls, contributing to their cooling. The thermal boundary layer grows for the two cases of mixed convection faster in the upper part than in the bottom, except for large eccentricities ($\varepsilon > 0.2$) when a downward shift is applied.

- From the work performed and the results obtained, it can be deduced that a proper choice of a downward eccentricity level (>0.2) reduces the thermal stratification observed at a moderate distance from the entrance of the heating zone for the concentric case. The choice of the adequate eccentricity value depends on rheological and thermal properties of the fluid. For the presently studied case, a value of eccentricity $\varepsilon \approx 0.4$ seems appropriate as it satisfies the abovementioned condition while still being farther enough from the flow blockage limit of 0.6.

The alternative of applying an eccentricity between the cylinders which is proposed in the present study is justified by its easy integration in industrial processes. Nevertheless, it would be interesting to extend this study to other geometries (two concentric or off centered ellipses, an ellipse and a cylinder ...) in order to propose the most suitable geometry for offsetting the thermal stratification phenomenon observed in the concentric annular duct case. This can be of great importance in cases of industrial processes where uniform temperature distribution at the exit of the heat exchanger is a sensitive issue. This is particularly critical when thermal stratification can have a denaturing effect on the fluid, because of the temperature differences it induces in the flow, and is therefore highly undesirable.

**Figures captions**

**Fig. 1** Geometry of the heated duct.

**Fig. 2** Bipolar coordinate system.

**Fig. 3** Mesh of the studied geometry.

**Fig. 4** Effect of eccentricity ($\varepsilon$) and shear-thinning index ($n$) on the fully developed axial velocity profile used at the duct entrance (Re=40). (a) $\beta= 0$ : wide part of cross section ; (b) $\beta= \pi$ : narrow part of cross section.

**Fig. 5** Axial variation of $\overline{Nu}_1$ for a thermally developing flow (mixed convection) for various transverse meshes ($\alpha \times \beta$). Upward shift; $\varepsilon=0.2$ and $\varepsilon=0.7$; $r_1=0.5$; Re=38.30; Pr=873.02; Gr=985.95; $n=1.0$; Pn=1.96.

**Fig. 6** Development of the thermal boundary layer along the heated duct for $\beta= 0$ and $\beta=\pi$. Case of forced convection; $n=0.7$; Pn=0; Re=40.5; Pr=1410. (a) $\varepsilon=0$; (b) $\varepsilon=0.6$.

**Fig. 7** Effect of eccentricity $\varepsilon$ on outer wall temperature evolution. Case of forced convection; $n=0.7$; Pn=0; Re=40.5; Pr=1410; $T_e$=20°C ; (----narrow part of cross section, —— wide part of cross section).

**Fig. 8** Variation of the axial velocity along the duct for $\beta=0$ and $\beta=\pi$. Case of forced convection; $n=0.7$; Pn=8; Re=40.5; Pr=1410; $T_e$= 20°C. (a) $\varepsilon=0$; (b) $\varepsilon=0.6$.

**Fig. 9** Variation of outer wall temperature $T_2$ along the heating zone for $\varepsilon=0.6$ and $n=0.7$. Case of forced convection; Pn=8; Re=40.5; Pr=1410; $T_e$= 20°C; (----narrow part of cross section, —— wide part of cross section).

**Fig. 10** Effect of eccentricity ($\varepsilon$) change on $\overline{Nu}_2$ along the heating zone. Case of forced convection; $n=1.0$; Pn=0 ; Re=27 ; Pr=891.

**Fig. 11** Effect of shear-thinning index ($n$) change on $\overline{Nu}_2$ along the heating zone for $\varepsilon=0.2$. Case of forced convection; Pn=0 ; Re=27 ; Pr=891.

**Fig. 12** Effect of thermodependency ($K(T)$) on $\overline{Nu}_2$ along the heating zone. Case of forced convection; $n=1.0$; Pn=8; Re=27; Pr=891.

**Fig. 13** Structure of the azimuthal flow at $z=1.75 \cdot 10^{-3}$ for $n=0.7$; Pn=0; Re=40.5; Pr=1410; Gr= 7497.

**Fig. 14** Tangential velocity profile at $z=3.50 \cdot 10^{-3}$ for $n=0.7$; Pn=0; Re=40.5; Pr=1410; Gr= 7497.

**Fig. 15** Effect of $n$ on the tangential velocity profile for $\varepsilon=0$; $z=5.25 \cdot 10^{-3}$; $\beta=\pi/2$; Case of : mixed convection Pn=0 ; Re=40.5 ; Pr=1410 ; Gr= 7497.

**Fig. 16** Development of the thermal boundary layer along the duct for $\beta= 0$ and $\beta=\pi$ and different $\varepsilon$ values for $n=0.7$; Pn=8; Re=40.5; Pr=1410; Gr=7497.

**Fig. 17** Variation of the axial velocity along the duct for $\beta= 0$ and $\beta=\pi$ for different $\varepsilon$ values for $n=0.7$; Pn=8; Re=40.5; Pr=1410; Gr=7497; $T_e=20°C$.

**Fig. 18** Variation of $\overline{Nu}_2$ along the heating zone for $\varepsilon=0.2$; $n=1.0$; Pn=0; Re=27; Pr=891 ; Gr=4258.

**Fig. 19** Variation of outer wall temperature $T_2$ along the duct for different $\varepsilon$ values for $n=0.7$; Pn=8; Re=40.5; Pr=1410; Gr=7497; $T_e=20°C$ ; (---- Top, — bottom).

**Fig. 20** Effect of thermodependency ($K(T)$) on $Nu_2$ for $\beta=0$ and $\beta=\pi$; $n=1.0$; Re= 27; Pr= 891; Gr= 4258 ; --- Pn=0; ---- Pn=8.0. (a) Downward shift; (b) Upward shift.

**Table 1:** Variation of $W_{max}/W_m$ for $n=1$ and $\beta=0$ with eccentricity (values between brackets are for a more refined mesh); (*a*) present results; (*b*) results of Feldman *et al.* [5], p239.

| Eccentricity $\varepsilon$ | Radius ratio $r_1$ ($R_1/R_2$) | $W_{max}/W_m$ ($\beta=0$)$_{(a)}$ | $W_{max}/W_m$ ($\beta=0$)$_{(b)}$ | Err (%) |
|---|---|---|---|---|
| 0.9 | 0.5 | 2.364 | 2.324 (2.310) | 2.33 |
| 0.5 | 0.5 | 2.389 | 2.373 (2.372) | 0.71 |
| 0.0 | 0.5 | 1508 | 1.508 | 0.00 |
| $10^{-3}$ | 0.4 | 1.526 | 1.516 | 0.66 |
| 0.0 | 0.4 | 1.513 | 1.513 | 0.00 |
| 0.7 | 0.3 | 2.289 | 2.277 (2.274) | 0.66 |
| 0.0 | 0.3 | 1.522 | 1.522 | 0.00 |
| 0.9 | 0.1 | 2.155 | 2.152 (2.076) | 3.80 |
| 0.5 | 0.1 | 2.148 | 2.149 (2.148) | 0.00 |
| 0.0 | 0.1 | 1.567 | 1.567 | 0.00 |